\documentclass[11pt,titlepage]{article}

\usepackage{setspace} 
\usepackage{amsmath}
\usepackage{graphicx}
\usepackage[round,numbers,sort&compress]{natbib}

\makeatletter
\long\def\@makecaption#1#2{%
  \vskip\abovecaptionskip
  \sbox\@tempboxa{#1: #2}%
  \ifdim \wd\@tempboxa >\hsize
    #1: #2\par
  \else
    \global \@minipagefalse
    \hb@xt@\hsize{\box\@tempboxa\hfil}%
  \fi
  \vskip\belowcaptionskip}
\makeatother

\begin{document}

\begin{titlepage}

\noindent
\textbf{\large How Noise and Coupling Induce Bursting Action Potentials
in Pancreatic $\boldsymbol{\beta}$-cells}
\vspace{2cm}

\noindent
\textbf{Junghyo Jo,$^{\star}$ Hyuk Kang,$^{\star}$ Moo Young Choi,$^{\star \dagger}$
and Duk-Su Koh$^{\ddagger}$}
\vspace{1cm}

\noindent
$^{\star}$Department of Physics, Seoul National University, Seoul 151-747, Korea;
$^{\dagger}$Korea Institute for Advanced Study, Seoul 130-722, Korea;
and $^{\ddagger}$Department of Physics, Pohang University of Science and Technology,
Pohang 790-784, Korea
\vspace{1cm}

\noindent
E-mail: mychoi@snu.ac.kr

\noindent
Corresponding author's present address:
Department of Physics, Seoul National University, Seoul 151-747, Korea;
and Korea Institute for Advanced Study, Seoul 130-722, Korea
\end{titlepage}

\abstract{
Unlike isolated $\beta$-cells, which usually produce continuous spikes
or fast and irregular bursts, electrically coupled $\beta$-cells are
apt to exhibit robust bursting action potentials. We consider the noise
induced by thermal fluctuations as well as that by channel gating
stochasticity and examine its effects on the action potential behavior
of the $\beta$-cell model. It is observed numerically that such noise
in general helps single cells to produce a variety of electrical activities.
In addition, we also probe coupling via gap junctions between neighboring cells,
with heterogeneity induced by noise, to find that it enhances regular bursts.

}

\clearpage

\section*{Introduction}
Bursting action potentials, which are characterized by rapid
firing interspersed with quiescent periods in pancreatic
$\beta$-cells, play a central role in the secretion of insulin,
the hormone for glucose homeostasis. It has been reported that
isolated $\beta$-cells actually show continuous spikes or fast and
irregular bursts \citep{Falke,Kinard,Smith} while $\beta$-cells in a
cluster or in an intact islet produce regular bursting action
potentials \citep{Andreu,Dean,Sanchez-Andres,Valdeolmillos}.
As for the correlations between the electrical activity on the cell
membrane and insulin secretion \citep{Henquin}, the
robust bursts appear more effective in maintaining glucose homeostasis
than continuous spikes, since coupled $\beta$-cells can control insulin
release better than isolated $\beta$-cells \citep{Bosco,Halban,Pipeleers}.

However, the question as to whether bursting is an endogenous
property of individual $\beta$-cells or of a cluster still remains
to be answered, which has attracted a number of investigations.
Among proposed explanations is the channel-sharing hypothesis,
which postulates that current fluctuations arising from channel
gating stochasticity prevent single cells, originally capable of
bursting, from bursting, but when they are electrically coupled,
the perturbing effects are shared by neighbors and the regular
bursting is recovered \citep{Aguirre,Chay1,Sherman1}.
In contrast to this hypothesis of negative
effects of noise, recent research \citep{Lee,Longtin,Pei,Pikovsky,Zaikin}
has established that noise can play a constructive role
in many biological systems including $\beta$-cell bursting \citep{Vries1}.
The heterogeneity hypothesis, providing another explanation,
was also postulated by the same group.  According to it, when heterogeneous cells,
each of which produces continuous spikes or bursts depending upon
such cell parameters as the size, channel density, etc., are coupled,
those cells in the cluster exhibit more pronounced bursts.
This gives a useful insight into the functioning of heterogeneous
cell populations \citep{Smolen}.

In this study, we expand the concept of heterogeneity and probe
how such general heterogeneity enhances bursting.  It is proposed that noise
induces heterogeneity in otherwise homogeneous individual $\beta$-cells,
which in turn assists the $\beta$-cells to produce robust bursts
when they are coupled.
Existing studies have mostly focused on the synchronizing role of
coupling \citep{Sherman2,Vries3};
the slow dynamics, which has a period about 10 to 60 seconds,
is synchronized successfully between adjacent cells.
In contrast, we focus here on the fact that rapid firing in the
active phase of bursting is asynchronous between neighbors \citep{Sherman3}
and these fluctuating currents
through the gap junction act like noise, enhancing the robust
bursting action potential.
It is also presented that various action potentials
of single $\beta$-cells are embodied with optimal noise induced by thermal
fluctuations or by ionic channel gating stochasticity.
In particular, noise stimulates occasionally itself to produce 
{\it fast bursts} in a single cell.

There are four sections in this paper: In the second section
the mathematical model for $\beta$-cells is introduced and the
simulation method is described. The third section is devoted to
the effects of random noise in currents and of voltage-dependent
noise in single cells while the fourth section examines how coupling
between cells influences the electrical activity of a cell.
Finally, main results are summarized and discussed in the last section.

\section*{Model and Methods}
\subsection*{Mathematical model for a $\beta$-cell}
As the Hodgkin-Huxley model \citep{Hodgkin} describes the
electrical activity on the cell membrane with ion channels, a few
mathematical models for $\beta$-cells, based on the
electrophysiological data \citep{Ashcroft,Gopel,Rorsman} of the ion channels in
$\beta$-cells, have been proposed. Although there are simple
models using two-dimensional maps \citep{Vries2,Rulkov1,Rulkov2},
we consider the Sherman model, which allows direct physical
interpretation \citep{Vries1,Sherman4}.

The model is described by the current balance equation between
capacitive and ionic currents:
\begin{eqnarray}
C_M \frac{dV}{dt} &=& - I_{Ca}(V) - I_K(V,N) -  I_{K(ATP)}(V,P)\nonumber \\
&& - I_S(V,S),  \label{V}
\end{eqnarray}
where $C_M$ and $V$ denote the membrane capacitance and the
membrane potential, respectively. The activation variable $N$ and
the slow variable $S$ are governed by
\begin{eqnarray}\label{N}
\tau_N \frac{dN}{dt} & = & N_{\infty}(V) - N  \nonumber \\
\tau_{S} \frac{dS}{dt} & = & S_{\infty}(V) - S
\end{eqnarray}
with appropriate relaxation times $\tau_N$ and $\tau_S$, which are taken to be
constants for simplicity. 
The fraction $P$ of open K(ATP) channels
may also be regarded as a constant for the moment [see Eq.~\ref{P}].
Ionic currents here are fast voltage-dependent L-type Ca$^{2+}$ current $I_{Ca}$,
delayed-rectifier K$^+$ current $I_K$, ATP-blockable K$^+$ current
$I_{K(ATP)}$, and very slow inhibitory potassium current $I_S$:
\begin{eqnarray}
\label{I}
I_{Ca}(V) & = & g_{Ca} M_{\infty}(V)(V - V_{Ca}) \nonumber\\
I_K(V, N) & = & g_K N (V - V_K) \nonumber \\ 
I_{K(ATP)}(V, P) & = & g_{K(ATP)} P (V - V_K)  \\
I_S(V, S) & = & g_S S (V - V_K).\nonumber
\end{eqnarray}
$I_{Ca}$ and $I_K$ are responsible for generating action
potentials; $I_{Ca}$ is assumed to respond instantaneously to a
change in the membrane potential, whereas $I_K$ is governed by the
dynamics of the activation variable $N$ via Eq.~\ref {N}.
$I_{K(ATP)}$ is the background current with voltage-independent
conductance $g_{K(ATP)}$; this determines the plateau fraction,
i.e., the ratio of the active phase duration to the burst period.
For example, as $g_{K(ATP)}$ decreases under high glucose
concentration, there are only active phases without silent phases.
$I_S$ is a phenomenological current representing slow dynamics in
the bursting action potential.
This model thus assumes that single $\beta$-cells originally
contain the slow dynamics, which works just under the appropriate
condition.  Biological candidates for such slow dynamics include
slow free Ca$^{2+}$ dynamics \citep{Chay2} and ATP
metabolism \citep{Keizer}. Finally, $M_{\infty}$,
$N_{\infty}$, and $S_{\infty}$ of the voltage-dependent activation
are defined to be
\begin{equation}
X_{\infty}(V) = \frac{1}{1 + \exp{[(V_X - V) / \theta_X]}},
\end{equation}
where $X$ denotes $M$, $N$, or $S$.

This set of coupled nonlinear differential equations in
Eqs.~\ref{V} - \ref{I} has been analyzed in
detail \citep{Keener,Rinzel}. There it is noted
that $S$ responds on a much slower time scale than $V$ and $N$
because $\tau_S$ has the time scale of several seconds compared
with the milli-second time scale in firing. Then $S$ is regarded
just as a parameter, and the dynamics of the fast subsystem on the
two-dimensional phase space of $V$ and $N$ is analyzed. Furthermore,
after eliminating one degree of freedom by substituting $N_{\infty}$
to $N$, the whole behavior of this model may be analyzed
approximately with fast variable $V$ and slow variable $S$.

\subsection*{Numerical details}
Integration of differential equations including noise demands some
caution, and is commonly achieved via the Euler method. For better
efficiency, we employ the Euler method for integrating the noise
term, combined with the second-order Runge-Kutta method for other
terms. In order to be concrete, we consider the one-variable
problem
\begin{equation}
\frac{dx}{dt} = f(x) + \xi(t),
\end{equation}
where $f(x(t))$ is a (nonlinear) function of $x$, the variable of
concern, and $\xi(t)$ is the white noise with zero mean and
delta-function correlations
\begin{eqnarray}
\langle \xi(t)\rangle & = & 0, \nonumber \\
\label{GWN} \langle \xi(t) \xi(t')\rangle & = & 2 D\delta(t-t').
\end{eqnarray}
Taking the time step of size $\Delta t$, we obtain from the
equation of motion the value of $x$ at time $t+\Delta t$:
\begin{equation}
x(t+\Delta t) = x(t) + \frac{f(x(t)) + f(\bar{x})}{2} \Delta t +
\xi(t){\Delta t},
\end{equation}
where $\bar{x} \equiv x(t) + f(x(t)) \Delta t + \xi(t){\Delta t}$ \citep{Batrouni}.
Although there is no gurantee that this algorithm should converge in general,
it works fine here since the noise term does not depend on the variable $x$ \citep{Kloeden}.

The white noise $\xi$ of variance $D$ is produced by the gaussian random
numbers with the variance $\sigma^2$ determined by
\begin{equation}
\langle\xi(t)^2\rangle
   = \int_{-\infty}^{\infty} d\xi \frac{1}{\sqrt{2 \pi} \sigma}
e^{-\xi^2/2\sigma^2}  \xi^2 = \frac{2D}{\Delta t},
\end{equation}
where the Dirac delta function has been represented by ${\Delta t}^{-1}$ 
within the numerical accuracy.
We thus have the relation $\sigma =\sqrt{2 D/\Delta t}$.

In our simulations, we take $\Delta t = 1$\,ms,
which turns out to be small enough, and integrate the set of equations for current
balance.  This gives the time evolution of the action potential,
from which the power spectrum is computed through the use of the
fast Fourier transform technique.

\section*{Results and Discussion}
\subsection*{Noise effects}
Before explaining the coupling effects, we first probe the role of
noise, either the usual (additive) random noise or the
(multiplicative) voltage-dependent one.  Comparison of the effects
of such noise helps us to understand better the coupling effects.

\subsubsection*{Random noise}
Among many kinds of noise on the cell membrane, the simplest case
is the random noise, which may come from thermal
fluctuations (see below). When such random current fluctuations
are present on the membrane, the current balance equation in
Eq.~\ref{V} is generalized to
\begin{eqnarray}
\label{V2}
C_M \frac{dV}{dt} & = & - I_{ion}(V,N,S) - \xi(t),
\end{eqnarray}
where $I_{ion}$ represents all the ionic currents on the
right-hand side in Eq.~\ref{V}, and the noise current $\xi (t)$
satisfies Eq.~\ref{GWN} with the variance denoted by $D_{\xi}$.

\begin{figure}
   \begin{center}
      \includegraphics*[width=1.2\textwidth]{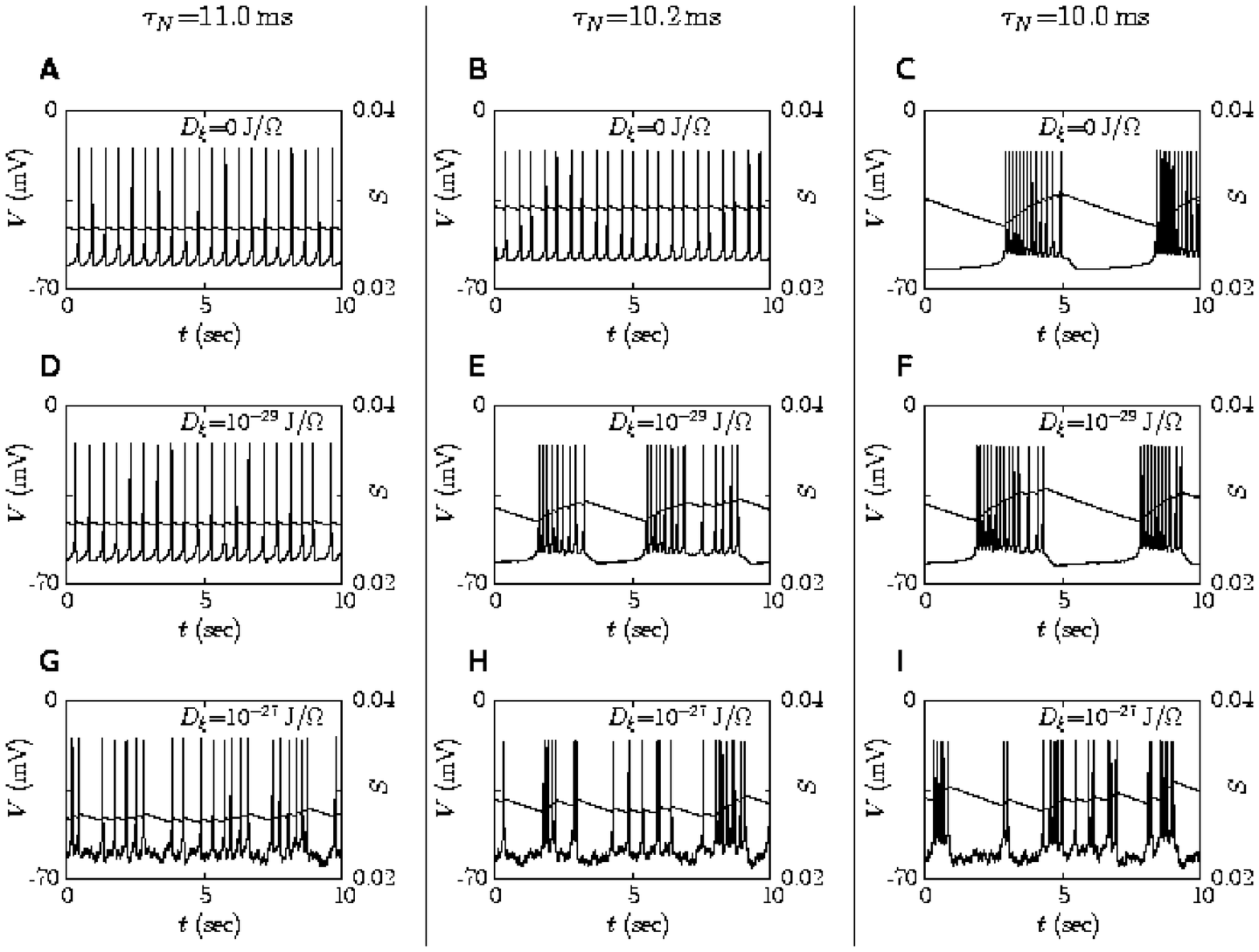}
      \caption{}
      \label{Fig:rd_noise}
   \end{center}
\end{figure}

Figure~\ref{Fig:rd_noise} exhibits the solution of the set of
coupled differential equations in Eqs.~\ref{N} 
and~\ref{V2} under various strengths of the random noise.
It is observed that single $\beta$-cells produce various
electrical activities according to the value of $\tau_N$ in
Eq.~\ref{N}, which lies in the narrow range 4 to
11\,ms depending on the membrane potential~(27).
When the time constant $\tau_N$ of delayed-rectifier K$^+$ channel
activity exceeds 11.0\,ms, the $\beta$-cell produces regular spiking
action potentials in Fig.~\ref{Fig:rd_noise} \textit{A},
while for $\tau_N$ below 10.0\,ms faster repolarization does not
allow enough time for the slow variable $S$ to decrease, yielding
bursting action potentials [see Fig.~\ref{Fig:rd_noise} \textit{C}].
In the intermediate regime of $\tau_N$=10.2\,ms,
Fig.~\ref{Fig:rd_noise} \textit{B} shows that spiking action potentials
are generated but the bursting property is resident.
As an appropriate amount of noise comes into play,
in particular, the regular spikes in Figs.~\ref{Fig:rd_noise} \textit{A}
and \textit{B} and bursts in \textit{C} change into fast bursts
in \textit{E}, irregular spikes in \textit{G}, or irregular bursts
in \textit{H} and \textit{I}.

To explain these phenomena, we note two thresholds of the slow
variable $S$: One is the upper threshold above which the membrane
potential is falling into the resting potential; the other is the
lower threshold above which the membrane potential begins to fire.
At the moment that fluctuations take negative values, they may
assist the repolarizing membrane potential to remain above the
lower threshold before the membrane repolarizes completely and then,
depolarizes slowly to the lower threshold. This induces
occasionally consecutive firing in Fig.~\ref{Fig:rd_noise} \textit{G}
or even fast bursts in Fig.~\ref{Fig:rd_noise} \textit{E} for the
$\beta$-cell in the critical parameter range, i.e.,
$\tau_N$=10.2\,ms.  Such consecutive firing raises the average
membrane potential for a while, compared with the case of regular
spikes. Hence the value of $S_{\infty}$ becomes large, and
consequently $S$ grows with the delay represented by the time
constant $\tau_S$. When it goes over the upper threshold, the
membrane potential returns to the resting potential. At the same
time, $S_{\infty}$ now becomes small and $S$ reduces to the lower
threshold. During this period of $S$ varying from the upper
threshold to the lower one, the membrane potential stays in the
silent phase. When $S$ comes to the lower threshold, the membrane
potential starts to depolarize and fire. Repetition of these
processes simply constitutes the fast bursts.
As the noise level is raised further, the slow variable $S$ may
start to increase before it reaches the lower threshold,
assisted by the fluctuations taking negative values.
Similarly it may start to decrease before it reaches the upper
threshold due to positive fluctuations.
In consequence, irregular bursts in Figs.~\ref{Fig:rd_noise} \textit{H}
and \textit{I} can thus be induced.
When fluctuations become sufficiently strong and dominant, such a role
of noise, turning on the slow dynamics of $S$, is concealed and
the membrane potential appears noisy.
Here it is notable that under optimal fluctuations, there exists
the critical parameter range in which the difference between the
upper and lower thresholds is small and the dramatic effect of fast bursts
is produced; similar results were obtained in a recent study \citep{Aguirre}.

\begin{figure}
   \begin{center}
      \includegraphics*[width=1.2\textwidth]{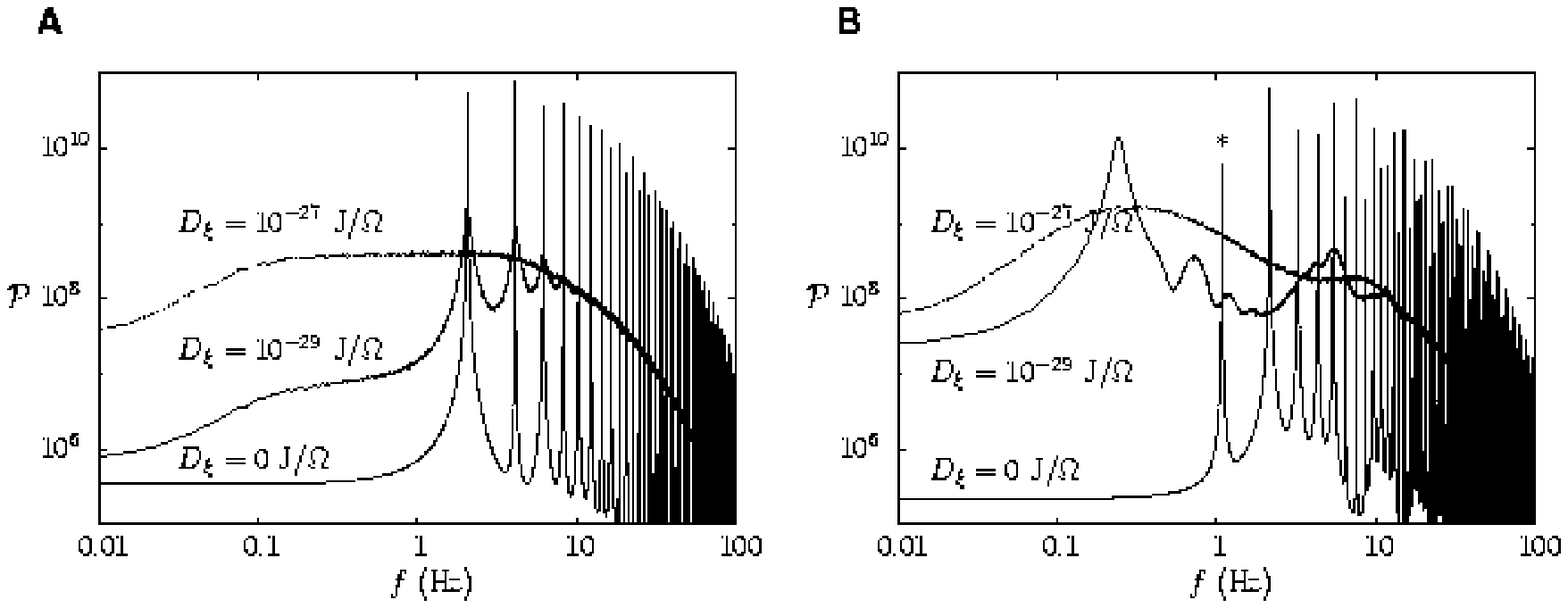}
      \caption{}
      \label{Fig:rd_power}
   \end{center}
\end{figure}

It is revealing to examine the power spectra of the obtained
action potentials, computed through the use of the fast Fourier
transform technique for various noise levels and displayed in
Fig.~\ref{Fig:rd_power}.
In particular, Figure~\ref{Fig:rd_power} \textit{B} manifests
that the regular spiking action potential of frequency 2\,Hz
in the absence of noise has changed into fast bursts containing
oscillations of 0.2\,Hz and 5\,Hz at moderate noise levels.

\begin{figure}
   \begin{center}
      \includegraphics*[width=0.6\textwidth]{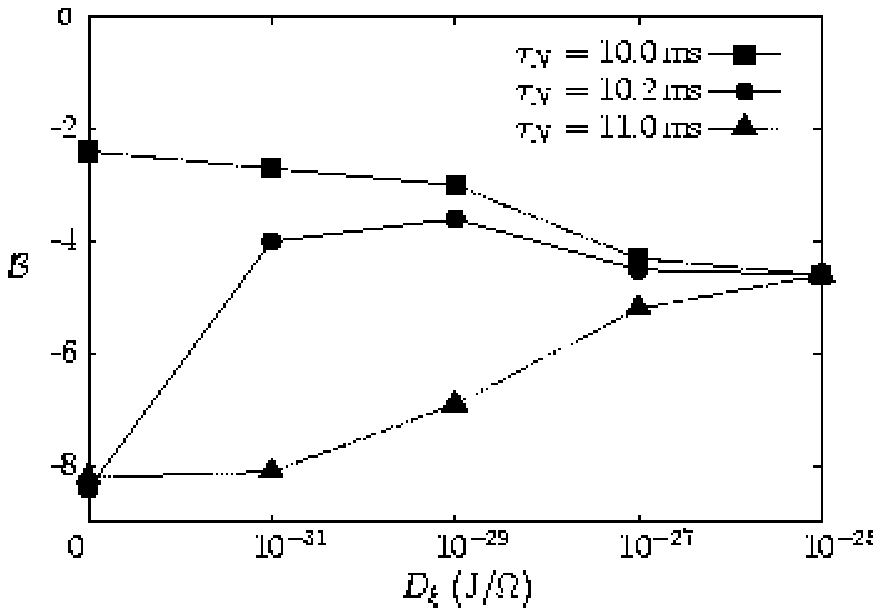}
      \caption{}
      \label{Fig:rd_summary}
   \end{center}
\end{figure}

To characterize the positive/negative role of noise in bursting,
we define the bursting tendency according to
$\cal{B} \equiv \log [\cal{P}($$f$$_B)/\cal{P}($$0)]$,
where $\cal{P}($$f_B)$ is the power spectrum at the bursting frequency $f_B$
and $\cal{P}($$0)$ is the background intensity at 0\,Hz.
Figure~\ref{Fig:rd_summary} shows the behavior of the bursting tendency $\cal{B}$
with the noise level, manifesting the noise effects on bursting.

Finally, one may ask whether thermal fluctuations known to
generate white noise are enough to induce the fast bursts,
irregular bursts or spikes, observed in our simulations.  In
simulations, the variance $D_{\xi}$ is taken in the range
$10^{-29}$\,J/$\Omega$ $\sim$ $10^{-27}$\,J/$\Omega$.  In reality,
noise currents due to thermal fluctuations can be estimated via
the fluctuation-dissipation theorem: $D_{\xi} = {k_B T}/{R}$. This
gives $D_{\xi} \sim 10^{-29}$\,J/$\Omega$
when $R$ is taken to be a few giga ohms (G$\Omega$) or less.
Accordingly, thermal fluctuations alone may not be enough
to induce irregular spikes or bursts.
Nevertheless, it appears possible that thermal fluctuations
actually expedite the emergence of fast bursts when the cell lies
in the critical parameter regime.

\subsubsection*{Voltage-dependent noise}
\begin{figure}
   \begin{center}
      \includegraphics*[width=1.2\textwidth]{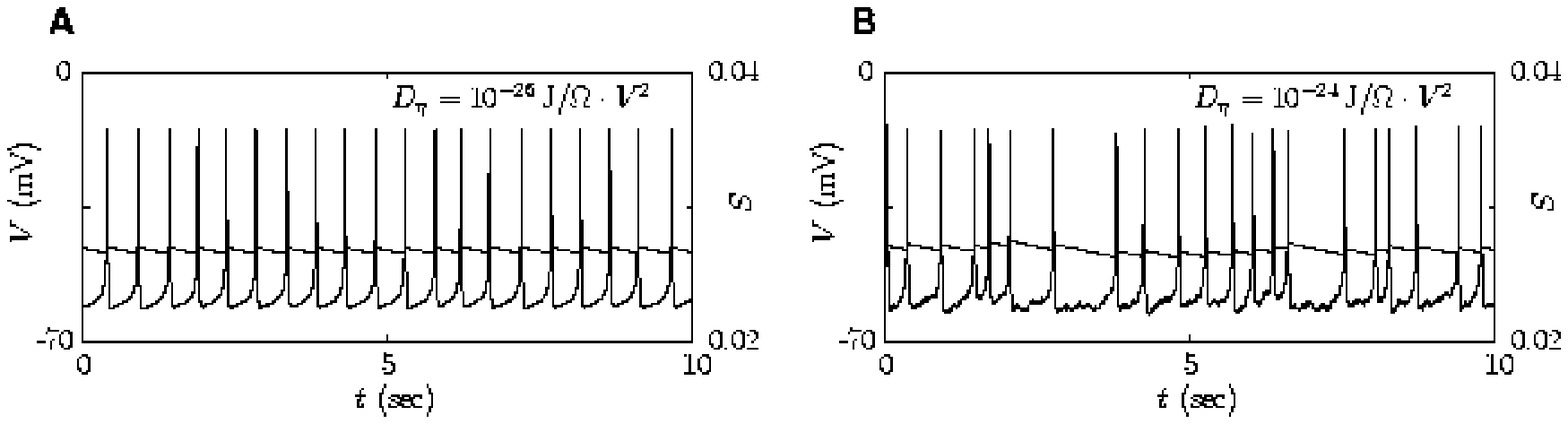}
      \caption{}
      \label{Fig:vd_noise}
   \end{center}
\end{figure}

As another simple type of noise, one can consider the voltage-dependent
fluctuations, which are closely related to the channel gating
stochasticity (see below). In the presence of such multiplicative noise,
the current balance condition in Eq.
\ref{V} takes the form
\begin{eqnarray}
\label{V3}
C_M \frac{dV}{dt} & = & - I_{ion}(V,N,S) -\eta(t)(V-V_K),
\end{eqnarray}
where $I_{ion}$ also represents all the ionic currents in
Eq.~\ref{V}, and $\eta(t)$ is the Gaussian white noise, again
satisfying Eq.~\ref{GWN} with variance $D_{\eta}$. Solving
numerically the coupled differential equations given by
Eqs.~\ref{N} and~\ref{V3} at various noise levels with
$\tau_N$ set equal to 11\,ms, we obtain the results, which are
illustrated in Fig.~\ref{Fig:vd_noise}. Note the overall similarity
to the case of random (additive) noise shown in
Figs.~\ref{Fig:rd_noise} \textit{D} and \textit{G}.

When the voltage-dependent noise stimulates the cell membrane,
irregular spikes arise, similarly to the case of random noise, if
its amplitude multiplied by the voltage difference $(V{-}V_K)$ is
comparable to the amplitude of random noise. In fact,
voltage-dependent noise may be regarded simply as the noise
weighted more in the active phase of the membrane potential than
in the silent phase. When taking negative values, therefore,
fluctuations boost firing more effectively in the active phase and
contribute less to the erratic evolution of the resting potential
in the silent phase.
Such voltage-dependent (multiplicative) noise may arise from ion
channel gating stochasticity, since currents through channels
depend upon the membrane potential difference. If the number of
channels is sufficiently large, the channel stochasticity can be
described by a Langevin equation \citep{DeFelice,Fox1,Fox2}.
Specifically, the stochasticity of K(ATP) channels has been
considered \citep{Vries1}. In the expression for the
ATP-dependent $K^+$ current, $I_{K(ATP)} = g_{K(ATP)} P (V - V_K)$,
the opening ratio $P$, which is no more constant, evolves
according to
\begin{eqnarray}
\label{P} \frac{dP}{dt}  =  \frac{\gamma_1}{\tau_{P}} (1 - {P}) -
\frac{\gamma_2}{\tau_{P}} P + \bar{\xi} (t),
\end{eqnarray}
where ${\gamma_1}/{\tau_{P}}$ and ${\gamma_2}/{\tau_{P}}$
represent the rates for a closed channel to switch to the open state
and vice versa, respectively.
Note that $\gamma_1$ and $\gamma_2$ thus determine the equilibrium ratio
between the open state and the closed one.
Fluctuations in the opening ratio are described by the Gaussian
white noise $\bar{\xi}(t)$ satisfying Eq.~\ref{GWN} with the variance
\begin{equation}
D_{\bar{\xi}} = \frac{\gamma_1 (1 - P) + \gamma_2 P}{2 \tau_{P} N_{K(ATP)}}
\approx \frac{\gamma_1 \gamma_2}{\tau_{P} N_{K(ATP)} (\gamma_1 +
\gamma_2)},
\label{Dbarxi}
\end{equation}
where $N_{K(ATP)}$ is the total number of ATP-dependent $K^+$
channels in a $\beta$-cell \citep{Fox1}.

\begin{figure}
   \begin{center}
      \includegraphics*[width=1.2\textwidth]{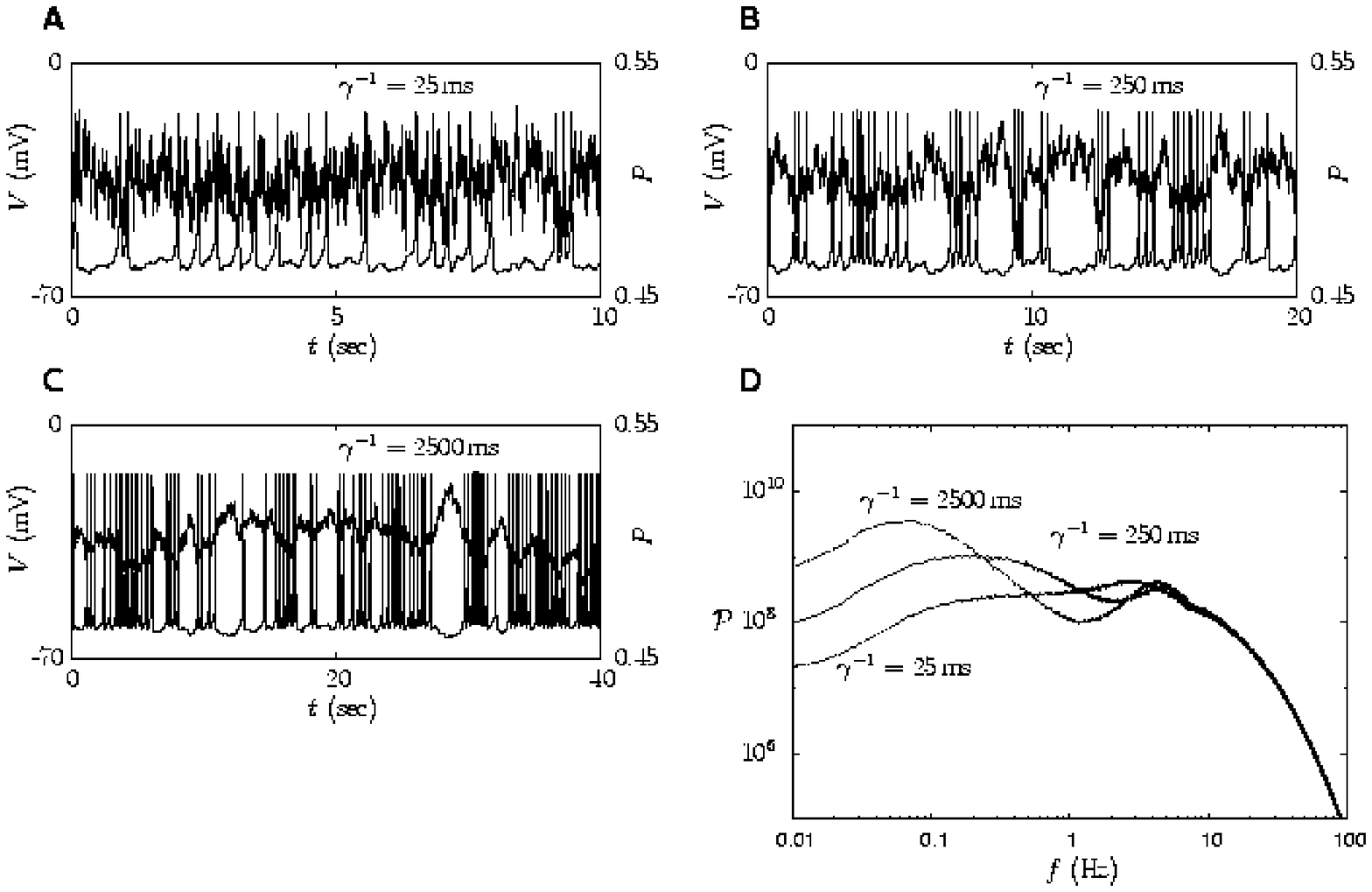}
      \caption{}
      \label{Fig:corr}
   \end{center}
\end{figure}

\begin{figure}
   \begin{center}
      \includegraphics*[width=1.2\textwidth]{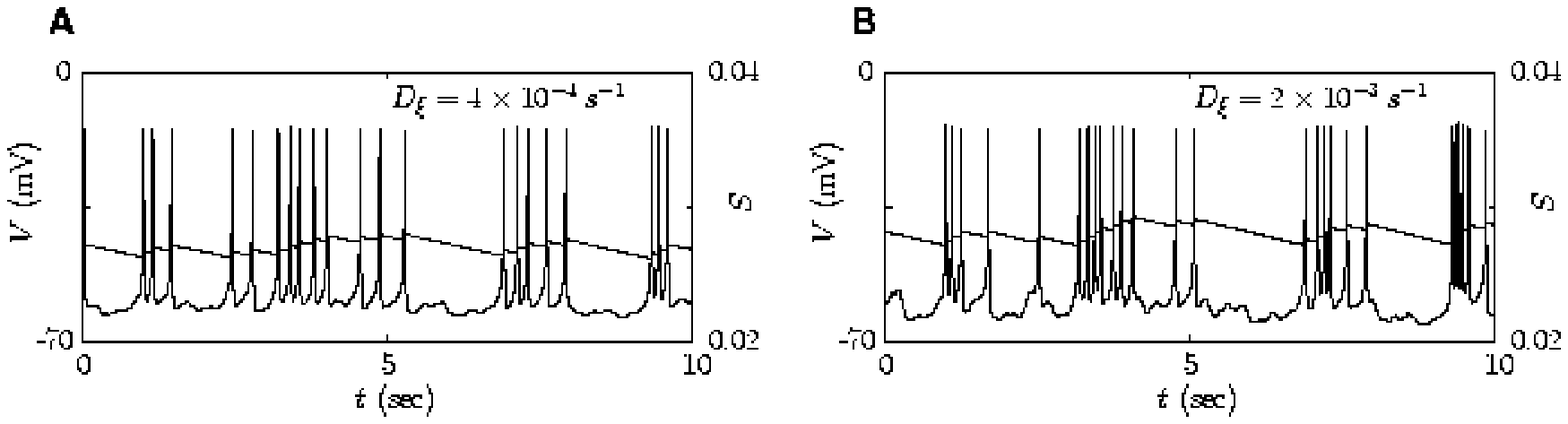}
      \caption{}
      \label{Fig:co_noise}
   \end{center}
\end{figure}

Solving Eq.~\ref{P}, we obtain that $P$ fluctuates around the equilibrium
value $P_0$, taken to be $0.5$ in our simulations: $P(t) = P_0 + \bar{\eta}(t)$.
Here $\bar{\eta}(t)$ is colored noise, characterized by the variance
\begin{eqnarray}
\langle\bar{\eta}(t)\bar{\eta}(t')\rangle = D_{\bar{\eta}}
[{\gamma}e^{-\gamma |t-t'|} - {\gamma}e^{-\gamma (t +t')}]
\end{eqnarray}
with $\gamma \equiv (\gamma_1 + \gamma_2)/\tau_p$ and
$D_{\bar{\eta}} \equiv D_{\bar{\xi}}/\gamma^2$ (see Appendix for details).
Note that the firing time scale is comparable to the correlation time $\gamma^{-1}$
of the noise $\bar{\eta}(t)$ (see Fig.~\ref{Fig:corr}).
Consequently this colored noise is more effective to induce several consecutive firings,
which resemble irregular burst, than the white noise.
In particular, the modules of several spikes are observed to become longer
as the correlation time $\gamma^{-1}$ is increased.
Figure~\ref{Fig:co_noise} shows the behaviors in the presence of the
channel-gating noise $\bar{\xi}(t)$ for two different channel numbers.
In this case of multiplicative colored noise, modules of spikes arise
more efficiently than in the case of mutiplicative white noise shown in Fig.~\ref{Fig:vd_noise}.
Further, it is also found that stronger gating fluctuations from less channels
($N_{K(ATP)}$=500) in Fig.~\ref{Fig:co_noise} \textit{B} give rise to modules
of more rapid spikes, compared with the case $N_{K(ATP)}$=2500 in Fig.~\ref{Fig:co_noise} \textit{A}.

Similar results can be obtained with fluctuations in the
Ca$^{2+}$ channels and in the delayed-rectifier K$^+$ channels
although they act somewhat differently from 
the fluctuations in the ATP-blockable K$^+$ channels (data not shown).

It is thus concluded that noise generates diverse firing patterns in single $\beta$-cells.
In a real (physiological) islet, however, $\beta$-cells are not isolated but coupled with each other,
making it desirable to consider coupled $\beta$-cells and to investigate effects of noise together with
those of coupling.  This will be the subject of the next section.

\subsection*{Coupling effects}
We consider two cells coupled with each other via a
gap junction.  With the coupling incorporated, Eq.~\ref{V} is
extended to the coupled equations:
\begin{eqnarray}
\label{V4}
C_{M} \frac{dV_1}{dt}&=& - I_{ion}(V_1,N_1,S_1,P_1) -g_C (V_1 - V_2) \nonumber \\
C_{M} \frac{dV_2}{dt}&=& - I_{ion}(V_2,N_2,S_2,P_2) -g_C (V_2 -
V_1),
\end{eqnarray}
where the subscripts $1$ and $2$ are the cell indices, $I_{ion}$
again denotes all the ionic currents, and $g_C$ is the coupling
conductance. Note that the heterogeneity between both cells is
accommodated in the K(ATP) channel opening ratio $P$. Namely, the
noise associated with channel gating stochasticity induces
continuously heterogeneity between the cells.

\begin{figure}
   \begin{center}
      \includegraphics*[width=1.2\textwidth]{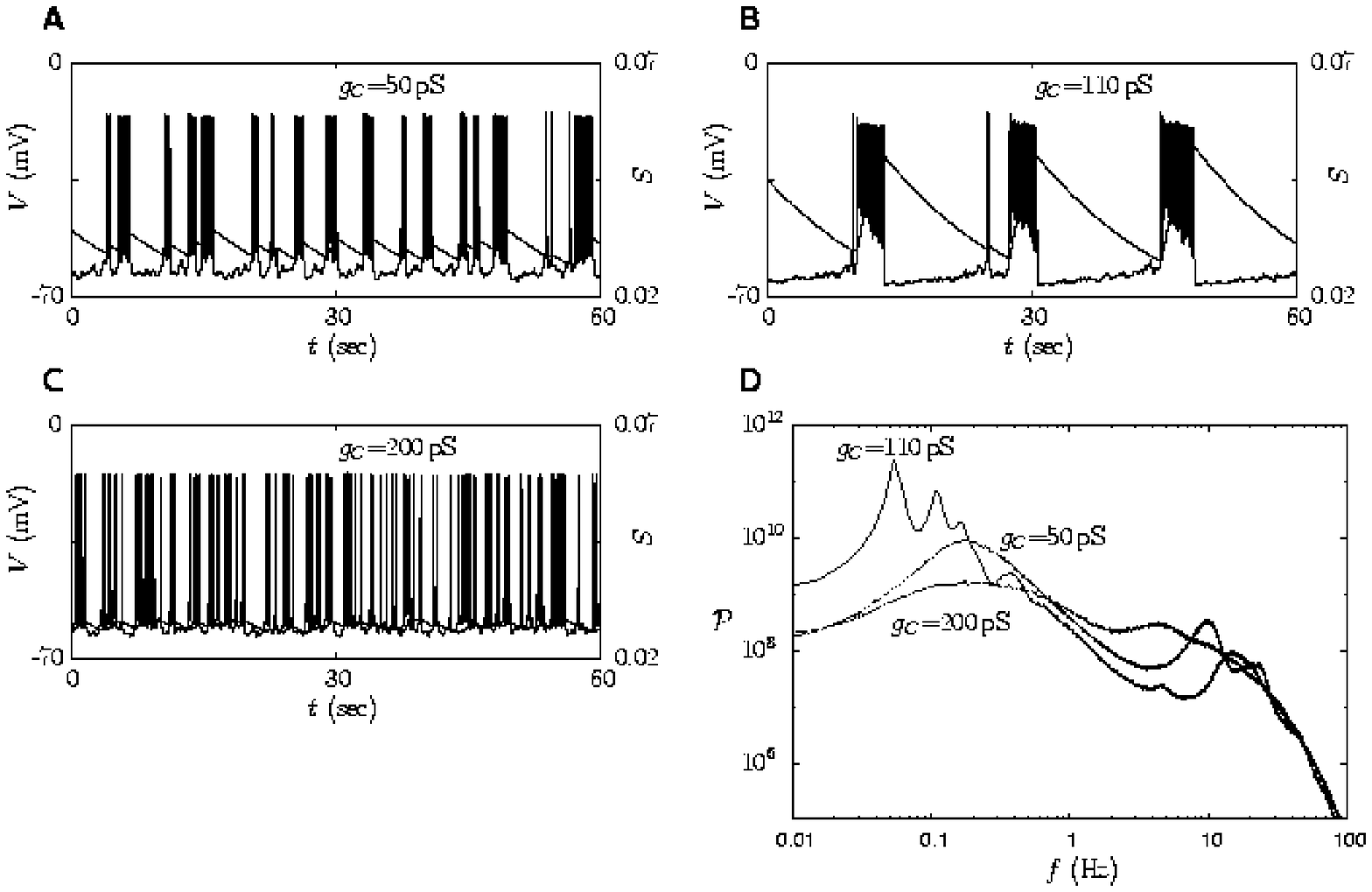}
      \caption{}
      \label{Fig:couple}
   \end{center}
\end{figure}

We thus have eight coupled differential equations, which consist
of Eqs.~\ref{N} and~\ref{P} for each cell and Eq.~\ref{V4},
for eight variables ($V, N, S$, and $P$ for each cell).
Integration of these coupled equations yields the results displayed
in Fig.~\ref{Fig:couple}, for the channel-gating noise of variance
$D_{\bar{\xi}}= 4 \times 10^{-4}$\,s$^{-1}$ given by Eq.~\ref{Dbarxi}
and for three values of the coupling conductance:
$g_C = 50$\,pS, $110$\,pS, and $200$\,pS.
Revealed is the optimal coupling strength for longer bursting periods:
While weak coupling is not enough to couple individual cells and
to generate consecutive firing, too strong coupling tends to make the
cluster behave as a single large cell \citep{Sherman2}.

\begin{figure}
   \begin{center}
      \includegraphics*[width=0.6\textwidth]{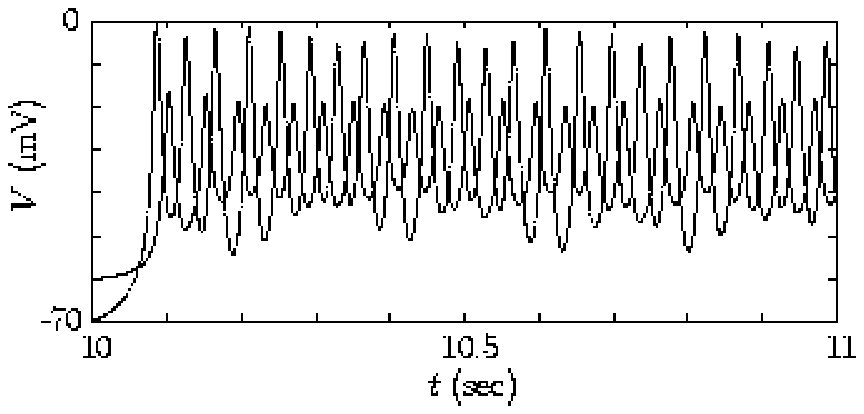}
      \caption{}
      \label{Fig:phase}
   \end{center}
\end{figure}

Robust bursts emerge as a consequence of the competition
between heterogeneity and coupling \citep{Vries3}.
On one hand, the coupling term in Eq.~\ref{V4} helps the two cells
to act synchronously; on the other hand, it also plays the role of
stimulating noise, which acts strongly on the two cells
with asynchronous phases.
The perfect asynchrony results from the harmony of coupling
to be similar and heterogeneity to be different (see Fig.~\ref{Fig:phase}).
Namely, the coupling currents between asynchronous neighboring cells give
rise to consecutive firing; this in turn increases the upper
threshold of the slow variable $S$ above which firing disappears.
As $S$ grows up toward the increased upper threshold, it takes
longer to reduce down to the lower threshold.
This larger rising and falling divides more clearly the active and silent
phases in the membrane potential, and accordingly induces
robust bursting action potentials with periods longer than 20\,s.
Note that in the absence of coupling we have not been able to observe
bursting periods longer than 10\,s
(see Figs.~\ref{Fig:rd_noise}-\ref{Fig:co_noise})
(Parameter values different from those in Table~I may
yield bursting periods somewhat longer than 10\,s even in a single cell.
In this case, the coupling gives rise to robust bursting of even longer
periods, say, 30\,s, still demonstrating its crucial role in generating
regular bursts.)

In the two-cell model here the optimal value of the coupling conductance
is observed to be $g_C = 110$\,pS. As the number of cells is increased,
however, more heterogeneity is introduced, which should be matched by
stronger coupling to generate robust bursts with longer periods.
Although the detailed investigation is beyond our computing capacity,
we have performed multi-cell simulations, which indeed confirms such
an increase of the optimal coupling conductance.
For example, the optimal conductance in the system of 1000 cells
turns out to be 100 to 300\,pS (data not shown),
which coincides with experimental results of the gap junctional conductance
\citep{Perezarmendariz}.

\begin{figure}
   \begin{center}
      \includegraphics*[width=0.6\textwidth]{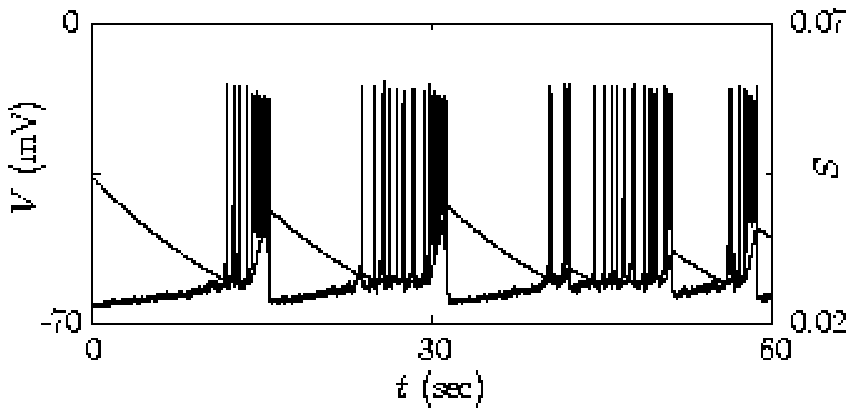}
      \caption{}
      \label{Fig:couple_vd}
   \end{center}
\end{figure}

These features of the coupled cells do not change much in the presence of
the voltage-dependent noise instead of the channel-gating noise,
except that the channel-gating noise is more efficient for robust bursting
than the voltage-dependent one, as shown in Fig.~\ref{Fig:couple_vd} for
$D_{\eta}= 10^{-24}$\,J/$\Omega\cdot$V$^2$ and $D_{\bar{\xi}}=0$.
Note also that the coupled cells depicted in Figs.~\ref{Fig:couple} and
~\ref{Fig:couple_vd} do not burst in the absence of noise-induced heterogeneity.

Recall that in the emergence of robust bursts, the asynchrony from the
heterogeneity induced by noise plays an important role,
which has also been addressed in a very recent study \citep{Pedersen}.
Similar to such noise-induced heterogeneity,
the cell-to-cell heterogeneity associated with variations of the cell
parameters among the cells is also expected to play for robust bursts \citep{Smolen}.
To check this, we allowed variations of the membrane capacitance $C_M$
related to the cell size as well as of the channel conductance $g_{K(ATP)}$
and examine the resulting behavior:
Shown in Fig.~\ref{Fig:hetero} \textit{A} and \textit{B}
are bursts generated in the case of 20\,\% variation
of $C_M$ (5.0\,pF, 6.3\,pF) and in the case of 10\,\% variation of $g_{K(ATP)}$
(1000\,pS, 1100\,pS), respectively.
Specifically, a spiking cell (with $C_M =6.3$\,pF) is coupled with a bursting cell
(with $C_M =5.0$\,pF) in Fig.~\ref{Fig:hetero} \textit{A},
which results in that both cells are bursting synchronously with
a longer bursting period than that of a single cell (5.0\,pF).
In Fig.~\ref{Fig:hetero} \textit{B}, on the other hand, two spiking cells
(with $g_{K(ATP)}=1000$\,pS and $1100$\,pS) are coupled with each other,
and both are bursting.
Therefore heterogeneity is in general important for bursting in coupled cells,
no matter whether it is cell-to-cell heterogeneity or induced by noise.

\section*{Conclusions}
We have probed whether noise and coupling serve as an appropriate
stimulus for inducing the bursting action potential in pancreatic
$\beta$-cells, and found that they effectively call into action
the inherent slow dynamics in individual cells. Fast bursts,
irregular spikes or bursts in single $\beta$-cells have been
observed as the results of the noise effects.  In particular the
emergence of regular bursts assisted by an appropriate amount of
noise [see Figs.~\ref{Fig:rd_noise} \textit{E} and~\ref{Fig:rd_power} \textit{B}]
is reminiscent of {\it coherence resonance} \citep{Lee,Longtin,Pei,Pikovsky,Zaikin}.
In view of physiology, the consecutive firing induced by
fluctuations gives rise to relative depolarization for a while,
which is followed by the activation of the slow potassium channel
lasting until the slow variable reaches the upper threshold.  At
this time the slow $K^+$ channel opens fully, and the outflux of
cytosolic potassium ions gets very large, thus hindering
depolarization.  Accordingly, the membrane potential is compelled
to stay in the silent phase, and the slow $K^+$ channel in turn
starts to be inactivated. In consequence, the membrane can become
depolarized as the outflux of $K^+$ ions reduces. Finally, firing
occurs again, and consecutive firing also happens by the help of
appropriate stimulation.  As candidates for the stimulus, both the
(additive) random noise coming from fluctuating currents and the
(multiplicative) voltage-dependent noise from the channel gating
stochasticity have been considered.

In particular, coupling between cells has turned out essential for
attaining regular bursts with longer periods compared with the
fast bursts. The coupling term, proportional to the potential
difference between two cells, operates in a similar manner to the
voltage-dependent noise: It increases with the potential
difference and thus becomes large for the cells in active phases,
stimulating the cells like noise. On the other hand, it is small
for perfectly synchronized cells in silent phases.  The coupling
also increases the upper threshold of $S$ and induces robust
regular bursts.

In the analysis, the heterogeneity has been found to play an important role
in inducing strong fluctuations during active phases, which may cause robust bursts.
Namely, bursting in general results from the interplay of
coupling and heterogeneity.
This allows us to interpret the fact that large cell clusters (up to
the critical size) show more regular bursts \citep{Vries1,Sherman2}:
Assuming a cubic islet, we have considered $\beta$-cells
arranged into an $L^3$ cube, under free boundary conditions.
Adopting physiological gap junction conductance, $g_C=200$\,pS~(42),
we have found that the bursting period and duration first increases with
the size $L$ but tends to saturate beyond $L=5$ (data not shown).
Such saturation behavior may be explained as follows:
Via the coupling through gap junctions, the number of nearest neighbors
in the three-dimensional space is limited, e.g., to six or so;
this suggests that the cluster above some critical size
can get no more advantage of the heterogeneity from
neighboring cells through given coupling strength.

The Langerhans islet, however, consists of several endocrine cells
in addition to $\beta$-cells. Other endocrine cells in an islet
have been studied recently \citep{Kanno,Nadal},
and it will be of interest to study the coupling effects between
originally different $\alpha$-, $\beta$-, and $\delta$-cells,
coupled via hormones or neurotransmitters \citep{Moriyama}.
This might give a clue to understanding the size of a Langerhans
islet in the pancreas, which is left for further study.

\section*{Appendix}
Equation~\ref{P} can be solved to give the time evolution of the
opening ratio $P$:
\begin{equation}
P(t) = P_0 + [ P(0) - P_0 ]  e^{-\gamma t}  + \int_{0}^{t}
e^{-\gamma (t-t')} \bar{\xi}(t') dt' \nonumber
\end{equation}
with $P_0 \equiv \gamma_1 / (\gamma_1 + \gamma_2 )$ and
$\gamma \equiv (\gamma_1 + \gamma_2)/\tau_p$,
where $P(0)$ is the initial value of $P$.
After sufficiently long time, we thus have $P$
fluctuating around the equilibrium ratio $P_0$: $P(t) = P_0 +
\bar{\eta}(t)$, where the noise $\bar{\eta}(t)$ is given by
\begin{eqnarray}
\bar{\eta}(t) \equiv \int_{0}^{t} e^{-\gamma (t-t')} \bar{\xi}(t') dt'.
\nonumber
\end{eqnarray}
From the above definition of the noise $\bar{\eta}(t)$, it is
straightforward to derive its characteristics:
\begin{eqnarray*}
\langle\bar{\eta}(t)\bar{\eta}(t')\rangle
&=& \int_0^t d\tau e^{\gamma(\tau-t)} \int_0^{t'} d\tau' e^{\gamma(\tau' -t')}
    \langle \bar{\xi}(\tau) \bar{\xi}(\tau') \rangle \\
&=& 2D_{\bar{\xi}} e^{-\gamma (t+t')} \int_0^{\bar{t}} d\tau e^{2\gamma\tau},
\end{eqnarray*}
where we have used the relation
$\langle \bar{\xi}(\tau) \bar{\xi}(\tau') \rangle = 2D_{\bar{\xi}} \delta (\tau -\tau')$
and $\bar{t}$ denotes the smaller one between $t$ and $t'$.
We thus obtain the correlations of the noise $\bar{\eta}$ at different times
\begin{equation*}
\langle\bar{\eta}(t)\bar{\eta}(t')\rangle
=  D_{\bar{\eta}}
[{\gamma}e^{-\gamma |t-t'|} - {\gamma}e^{-\gamma (t +t')}]
\end{equation*}
with $D_{\bar{\eta}} \equiv D_{\bar{\xi}}/\gamma^2$,
which manifests the colored nature.

\vspace{2cm}
This work was supported in part by KOSEF through Grant
No. 01-2002-000-00285-0 and by the MOST (KOSEF) through
National Core Research Center for Systems Bio-Dynamics, as well as
by the BK21 Program.
Helpful reprints from the Laboratory of Biological Modeling
in NIDDK of NIH are also gratefully acknowledged.


\begin{thebibliography}{46}
\expandafter\ifx\csname natexlab\endcsname\relax\def\natexlab#1{#1}\fi

\bibitem[{Falke et~al.(1989)Falke, Gillis, Pressel, and Misler}]{Falke}
Falke, L.~C., K.~D. Gillis, D.~M. Pressel, and S.~Misler. 1989.
\newblock `Perforated patch recording' allows long-term monitoring of
  metabolite-induced electrical activity and voltage-dependent Ca$^{2+}$
  currents in pancreatic islet $\beta$-cells.
\newblock \emph{FEBS Lett.} 251:167--172.

\bibitem[{Kinard et~al.(1999)Kinard, de~Vries, Sherman, and Satin}]{Kinard}
Kinard, T.~A., G.~de~Vries, A.~Sherman, and L.~S. Satin. 1999.
\newblock Modulation of the bursting properties of single mouse pancreatic
  $\beta$-cells by artificial conductances.
\newblock \emph{Biophys. J.} 76:1423--1435.

\bibitem[{Smith et~al.(1990)Smith, Ashcroft, and Rorsman}]{Smith}
Smith, P.~A., F.~M. Ashcroft, and P.~Rorsman. 1990.
\newblock Simultaneous recordings of glucose dependent electrical activity and
  atp-regulated K$^+$-currents in isolated mouse pancreatic $\beta$-cells.
\newblock \emph{FEBS Lett.} 261:187--190.

\bibitem[{Andreu et~al.(1997)Andreu, Soria, and S\'anchez-Andr\'es}]{Andreu}
Andreu, E., B.~Soria, and J.~V. S\'anchez-Andr\'es. 1997.
\newblock Oscillation of gap junction electrical coupling in the mouse
  pancreatic islets of langerhans.
\newblock \emph{J. Physiol.} 498:753--761.

\bibitem[{Dean and Matthews(1968)}]{Dean}
Dean, P.~M., and E.~K. Matthews. 1968.
\newblock Electrical activity in pancreatic islet cells.
\newblock \emph{Nature} 219:389--390.

\bibitem[{S\'anchez-Andr\'es et~al.(1995)S\'anchez-Andr\'es, Gomis, and
  Valdeolmillos}]{Sanchez-Andres}
S\'anchez-Andr\'es, J.~V., A.~Gomis, and M.~Valdeolmillos. 1995.
\newblock The electrical activity of mouse pancreatic $\beta$-cells recorded in
  vivo shows glucose-dependent oscillations.
\newblock \emph{J. Physiol.} 486:223--228.

\bibitem[{Valdeolmillos et~al.(1996)Valdeolmillos, Gomis, and
  S\'anchez-Andr\'es}]{Valdeolmillos}
Valdeolmillos, M., A.~Gomis, and J.~V. S\'anchez-Andr\'es. 1996.
\newblock In vivo synchronous membrane potential oscillations in mouse
  pancreatic $\beta$-cells: lack of co-ordination between islets.
\newblock \emph{J. Physiol.} 493:9--18.

\bibitem[{Henquin and Meissner(1984)}]{Henquin}
Henquin, J.~C., and H.~P. Meissner. 1984.
\newblock Significance of ionic fluxes and changes in membrane potential for
  stimulus-secretion coupling in pancreatic $\beta$-cells.
\newblock \emph{Experientia} 40:1043--1052.

\bibitem[{Bosco et~al.(1989)Bosco, Orci, and Meda}]{Bosco}
Bosco, D., L.~Orci, and P.~Meda. 1989.
\newblock Homologous but not heterologous contact increases the insulin
  secretion of individual pancreatic $\beta$-cells.
\newblock \emph{Exp. Cell Res.} 184:72--80.

\bibitem[{Halban et~al.(1982)Halban, Wollheim, Blondel, Meda, Niesor, and
  Mintz}]{Halban}
Halban, P.~A., C.~B. Wollheim, B.~Blondel, P.~Meda, E.~N. Niesor, and D.~H.
  Mintz. 1982.
\newblock The possible importance of contact between pancreatic islet cells for
  the control of insulin release.
\newblock \emph{Endocrinology} 111:86--94.

\bibitem[{Pipeleers et~al.(1982)Pipeleers, Veld, Maes, and Winkel}]{Pipeleers}
Pipeleers, D., P.~I. Veld, E.~Maes, and M.~V.~D. Winkel. 1982.
\newblock Glucose-induced insulin release depends on functional cooperation
  between islet cells.
\newblock \emph{Proc. Natl. Acad. Sci. USA} 79:7322--7325.

\bibitem[{Aguirre et~al.(2004)Aguirre, Mosekilde, and Sanju\'an}]{Aguirre}
Aguirre, J., E.~Mosekilde, and M.~A.~F. Sanju\'an. 2004.
\newblock Analysis of the noise-induced bursting-spiking transition in a
  pancreatic $\beta$-cell model.
\newblock \emph{Phys. Rev. E} 69:041910.

\bibitem[{Chay and Kang(1988)}]{Chay1}
Chay, T.~R., and H.~S. Kang. 1988.
\newblock Role of single-channel stochastic noise on bursting clusters of
  pancreatic $\beta$-cells.
\newblock \emph{Biophys. J.} 54:427--435.

\bibitem[{Sherman et~al.(1988)Sherman, Rinzel, and Keizer}]{Sherman1}
Sherman, A., J.~Rinzel, and J.~Keizer. 1988.
\newblock Emergence of organized bursting in clusters of pancreatic
  $\beta$-cells by channel sharing.
\newblock \emph{Biophys. J.} 54:411--425.

\bibitem[{Lee et~al.(1998)Lee, Neiman, and Kim}]{Lee}
Lee, S.~G., A.~Neiman, and S.~Kim. 1998.
\newblock Coherence resonance in a hodgkin-huxley neuron.
\newblock \emph{Phys. Rev. E} 57:3292--3297.

\bibitem[{Longtin(1997)}]{Longtin}
Longtin, A. 1997.
\newblock Autonomous stochastic resonance in bursting neurons.
\newblock \emph{Phys. Rev. E} 55:868--876.

\bibitem[{Pei et~al.(1996)Pei, Wilkens, and Moss}]{Pei}
Pei, X., L.~Wilkens, and F.~Moss. 1996.
\newblock Noise-mediated spike timing precision from aperiodic stimuli in an
  array of hodgkin-huxley-type neuron.
\newblock \emph{Phys. Rev. Lett.} 77:4679--4682.

\bibitem[{Pikovsky and Kurths(1997)}]{Pikovsky}
Pikovsky, A.~S., and J.~Kurths. 1997.
\newblock Coherence resonance in a noise-driven excitable system.
\newblock \emph{Phys. Rev. Lett.} 78:775--778.

\bibitem[{Zaikin et~al.(2003)Zaikin, Garc\'{\i}a-Ojalvo, B\'ascones, Ullner,
  and Kurths}]{Zaikin}
Zaikin, A., J.~Garc\'{\i}a-Ojalvo, R.~B\'ascones, E.~Ullner, and J.~Kurths.
  2003.
\newblock Doubly stochastic coherence via noise-induced symmetry in bistable
  neural models.
\newblock \emph{Phys. Rev. Lett.} 90:03061.

\bibitem[{de~Vries and Sherman(2000)}]{Vries1}
de~Vries, G., and A.~Sherman. 2000.
\newblock Channel sharing in pancreatic $\beta$-cells revisited: enhancement of
  emergent bursting by noise.
\newblock \emph{J. Theor. Biol.} 207:513--530.

\bibitem[{Smolen et~al.(1993)Smolen, Rinzel, and Sherman}]{Smolen}
Smolen, P., J.~Rinzel, and A.~Sherman. 1993.
\newblock Why pancreatic islets burst but single beta cells do not. the
  heterogeneity hypothesis.
\newblock \emph{Biophys. J.} 64:1668--1680.

\bibitem[{Sherman and Rinzel(1991)}]{Sherman2}
Sherman, A., and J.~Rinzel. 1991.
\newblock Model for synchronization of pancreatic $\beta$-cells by gap junction
  coupling.
\newblock \emph{Biophys. J.} 59:547--559.

\bibitem[{de~Vries and Sherman(2001)}]{Vries3}
de~Vries, G., and A.~Sherman. 2001.
\newblock From spikers to bursters via coupling: help from heterogeneity.
\newblock \emph{Bull. Math. Biol.} 63:371--391.

\bibitem[{Sherman and Rinzel(1992)}]{Sherman3}
Sherman, A., and J.~Rinzel. 1992.
\newblock Rhythmogenic effects of weak electrotonic coupling in neuronal
  models.
\newblock \emph{Proc. Natl. Acad. Sci. USA} 89:2471--2474.

\bibitem[{Hodgkin and Huxley(1952)}]{Hodgkin}
Hodgkin, A.~L., and A.~F. Huxley. 1952.
\newblock A quantitative description of membrane current and its application to
  conduction and excitation in nerve.
\newblock \emph{J. Physiol.} 117:500--544.

\bibitem[{Ashcroft and Rorsman(1989)}]{Ashcroft}
Ashcroft, F.~M., and P.~Rorsman. 1989.
\newblock Electrophysiology of the pancreatic $\beta$-cell.
\newblock \emph{Prog. Biophys. Mol. Biol.} 54:87--143.

\bibitem[{G\"opel et~al.(1999)G\"opel, Kanno, Barg, Galvanovskis, and
  Rorsman}]{Gopel}
G\"opel, S., T.~Kanno, S.~Barg, J.~Galvanovskis, and P.~Rorsman. 1999.
\newblock Voltage-gated and resting membrane currents recorded from
  $\beta$-cells in intact mouse pancreatic islets.
\newblock \emph{J. Physiol.} 521:717--728.

\bibitem[{Rorsman and Trube(1986)}]{Rorsman}
Rorsman, P., and G.~Trube. 1986.
\newblock Calcium and delayed potassium currents in mouse pancreatic
  $\beta$-cells under voltage-clamp conditions.
\newblock \emph{J. Physiol.} 374:531--550.

\bibitem[{de~Vries(2001)}]{Vries2}
de~Vries, G. 2001.
\newblock Bursting as an emergent phenomenon in coupled chaotic maps.
\newblock \emph{Phys. Rev. E} 64:051914.

\bibitem[{Rulkov(2001)}]{Rulkov1}
Rulkov, N.~F. 2001.
\newblock Regularization of synchronized chaotic bursts.
\newblock \emph{Phys. Rev. Lett.} 86:183--186.

\bibitem[{Rulkov(2002)}]{Rulkov2}
Rulkov, N.~F. 2002.
\newblock Modeling of spiking-bursting neural behavior using two-dimensional
  map.
\newblock \emph{Phys. Rev. E} 65:041922.

\bibitem[{Sherman(1996)}]{Sherman4}
Sherman, A. 1996.
\newblock Contributions of modeling to understanding stimulus-secretion
  coupling in pancreatic $\beta$-cells.
\newblock \emph{Am. J. Physiol.} 271:362--372.

\bibitem[{Chay and Keizer(1983)}]{Chay2}
Chay, T.~R., and J.~Keizer. 1983.
\newblock Minimal model for membrane oscillations in the pancreatic beta-cell.
\newblock \emph{Biophys. J.} 42:181--190.

\bibitem[{Keizer and Magnus(1989)}]{Keizer}
Keizer, J., and G.~Magnus. 1989.
\newblock ATP-sensitive potassium channel and bursting in the pancreatic beta
  cell. a theoretical study.
\newblock \emph{Biophys. J.} 56:229--242.

\bibitem[{Keener and Sneyd(1998)}]{Keener}
Keener, J., and J.~Sneyd. 1998.
\newblock Mathematical Physiology.
\newblock Springer-Verlag, New York, 188--215.

\bibitem[{Rinzel(1987)}]{Rinzel}
Rinzel, J. 1987.
\newblock A formal classification of bursting mechanisms in excitable systems.
\newblock \emph{In} Mathematical Topics in Population Biology, Morphogenesis,
  and Neurosciences, E.~Teramoto, and M.~Yamaguti, editors. Springer-Verlag,
  New York, 267--281.

\bibitem[{Batrouni et~al.(1985)Batrouni, Katz, Kronfeld, Lepage, Svetitsky, and
  Wilson}]{Batrouni}
Batrouni, G.~G., G.~R. Katz, A.~S. Kronfeld, G.~P. Lepage, B.~Svetitsky, and
  K.~G. Wilson. 1985.
\newblock Langevin simulations of lattice field theories.
\newblock \emph{Phys. Rev. D} 32:2736.

\bibitem[{Kloeden et~al.(1994)Kloeden, Platen, and Schurz}]{Kloeden}
Kloeden, P.~E., E.~Platen, and H.~Schurz. 1994.
\newblock Numerical Solution of SDE through Computer Experiments.
\newblock Springer-Verlag, Berlin, 151.

\bibitem[{DeFelice and Isaac(1992)}]{DeFelice}
DeFelice, L.~J., and A.~Isaac. 1992.
\newblock Chaotic states in a random world: relationship between the nonlinear
  differential equations of excitability and the stochastic properties of ion
  channels.
\newblock \emph{J. Stat. Phys.} 70:339--354.

\bibitem[{Fox and Lu(1994)}]{Fox1}
Fox, R.~F., and Y.~N. Lu. 1994.
\newblock Emergent collective behavior in large numbers of globally coupled
  independently stochastic ion channels.
\newblock \emph{Phys. Rev. E} 49:3421--3431.

\bibitem[{Fox(1997)}]{Fox2}
Fox, R.~F. 1997.
\newblock Stochastic versions of the hodgkin-huxley equations.
\newblock \emph{Biophys. J.} 72:2068--2074.

\bibitem[{P\'erez-Armendariz et~al.(1991)P\'erez-Armendariz, Roy, Sparay, and
  Bennett}]{Perezarmendariz}
P\'erez-Armendariz, M., C.~Roy, D.~C. Sparay, and M.~V.~L. Bennett. 1991.
\newblock Biophysical properties of gap junctions between freshly dispersed
  pairs of mouse pancreatic beta cells.
\newblock \emph{Biophys. J.} 59:76--92.

\bibitem[{Pedersen(2005)}]{Pedersen}
Pedersen, M.~G. 2005.
\newblock A comment on noise enhanced bursting in pancreatic $\beta$-cells.
\newblock \emph{J. Theor. Biol.} 235:1--3.

\bibitem[{Kanno et~al.(2002)Kanno, G\"opel, Rorsman, and Wakui}]{Kanno}
Kanno, T., S.~O. G\"opel, P.~Rorsman, and M.~Wakui. 2002.
\newblock Cellular function in multicellular system for hormone-secretion:
  electrophysiological aspect of studies on $\alpha$-, $\beta$- and
  $\delta$-cells of the pancreatic islet.
\newblock \emph{Neurosci. Res.} 42:79--90.

\bibitem[{Nadal et~al.(1999)Nadal, Quesada, and Soria}]{Nadal}
Nadal, A., I.~Quesada, and B.~Soria. 1999.
\newblock Homologous and heterologous asynchronicity between identified alpha-,
  beta- and delta-cells within intact islets of langerhans in the mouse.
\newblock \emph{J. Physiol.} 517:85--93.

\bibitem[{Moriyama and Hayashi(2003)}]{Moriyama}
Moriyama, Y., and M.~Hayashi. 2003.
\newblock Glutamate-mediated signaling in the islets of langerhans: a thread
  entangled.
\newblock \emph{TRENDS in Pharmacol. Sci.} 42:511--517.

\end{thebibliography}

\clearpage

\section*{Table}
\begin{table}[h]
\caption{Standard parameter values}
\begin{tabular}{cc}
\hline
$C_M = 6.3$\,pF & $g_{Ca} = 3000$\,pS  \\
$g_{K} = 4000$\,pS & $g_{K(ATP)} = 1000$\,pS \\
$g_{S} = 3000$\,pS & $g_C = 110$\,pS  \\
$V_{Ca} = 25$\,mV & $V_{K} = -75$\,mV   \\
$V_{M} = -20$\,mV & $\theta_{M} = 12$\,mV  \\
$V_{N} = -17$\,mV & $\theta_{N} = 5.6$\,mV  \\
$V_{S} = -22$\,mV & $\theta_{S} = 8.0$\,mV  \\
$\tau_N = 1.1\times 10^{-2}$\,s & $\tau_S = 20$\,s  \\
$\tau_P = 0.50$\,s & $N_{K(ATP)} = 2500$\, \\
$\gamma_1 = 1$ & $\gamma_2 = 1$  \\
\hline
\end{tabular}
\label{tab:table1}
\end{table}

\clearpage
\section*{Figure Legends}
\subsubsection*{Figure~\ref{Fig:rd_noise}.}
Action potential $V$ and slow channel activity
$S$ in single $\beta$-cells at the noise level $D_{\xi}=0, 10^{-29}$,
and $10^{-27}$\,J/$\Omega$ under several values of time constant
$\tau_N$ of delayed-rectifier K$^+$ channel activity $N$.  All
simulations have been performed under the standard parameter
values in Table~I except $\tau_N$, the values of
which are given above.

\subsubsection*{Figure~\ref{Fig:rd_power}.}
Power spectra of the action potentials for
the random noise levels in Fig.~\ref{Fig:rd_noise}. The time constant
$\tau_N$ of the activation variable $N$ is (\textit{A}) 11.0\,ms and
(\textit{B}) 10.2\,ms. Observed in the power spectra are main peaks
together with their harmonics. The peak at 1\,Hz, indicated by the asterisk
in (\textit{B}), reflects the tendency to form dimerization of spikes.
Each power spectrum has been obtained from the average over
1000 samples, each having a time sequence of $132$ seconds.

\subsubsection*{Figure~\ref{Fig:rd_summary}.}
Bursting tendency $\cal{B}$ of $\beta$-cells versus the noise level 
for several values of $\tau_N$, corresponding to different firing patterns
in the absence of noise.

\subsubsection*{Figure~\ref{Fig:vd_noise}.}
Action potential $V$ and slow channel activity $S$
in single $\beta$-cells at two values of the voltage-dependent noise.
Again parameter values in Table~I have been used.

\subsubsection*{Figure~\ref{Fig:corr}.}
Correlations between the action potential and
multiplicative colored noise due to channel-gating stochasticity.
The correlation time $\gamma^{-1}$ is taken to be (\textit{A}) 25\,ms,
(\textit{B}) 250\,ms, and (\textit{C}) 2500\,ms. Note that each figure
has a different time scale. Their corresponding power spectra are shown in (\textit{D}).
Parameter values in Table~I have been used except $\tau_P$.

\subsubsection*{Figure~\ref{Fig:co_noise}.}
Action potential $V$ and slow channel activity $S$
in single $\beta$-cells at two values of the channel gating stochasticity:
(\textit{A}) and (\textit{B}) correspond to the channel number
$N_{K(ATP)}= 2500$ and $500$, respectively.
Note that Figs.~\ref{Fig:corr} \textit{B} and \ref{Fig:co_noise} \textit{A}
represent the same sample path, but with different variables plotted.
Other parameter values have been taken from Table~I.

\subsubsection*{Figure~\ref{Fig:couple}.}
Figure~\ref{Fig:phase}. Enlarged view of the interval between 10\,s to 11\,s in
Fig.~\ref{Fig:couple} \textit{B}, disclosing the detailed behavior of the two
membrane potentials $V_1$ (\textit{solid line}) and $V_2$ (\textit{dashed line}).

\subsubsection*{Figure~\ref{Fig:couple_vd}.}
Bursting action potential induced
by cell coupling via the gap junction of conductance $g_C =110$\,pS
under the voltage-dependent noise of strength
$D_{\eta}=10^{-24}$\,J/$\Omega\cdot$V$^2$.
Parameter values in Table~I have been used.

\begin{figure}
   \begin{center}
      \includegraphics*[width=1.2\textwidth]{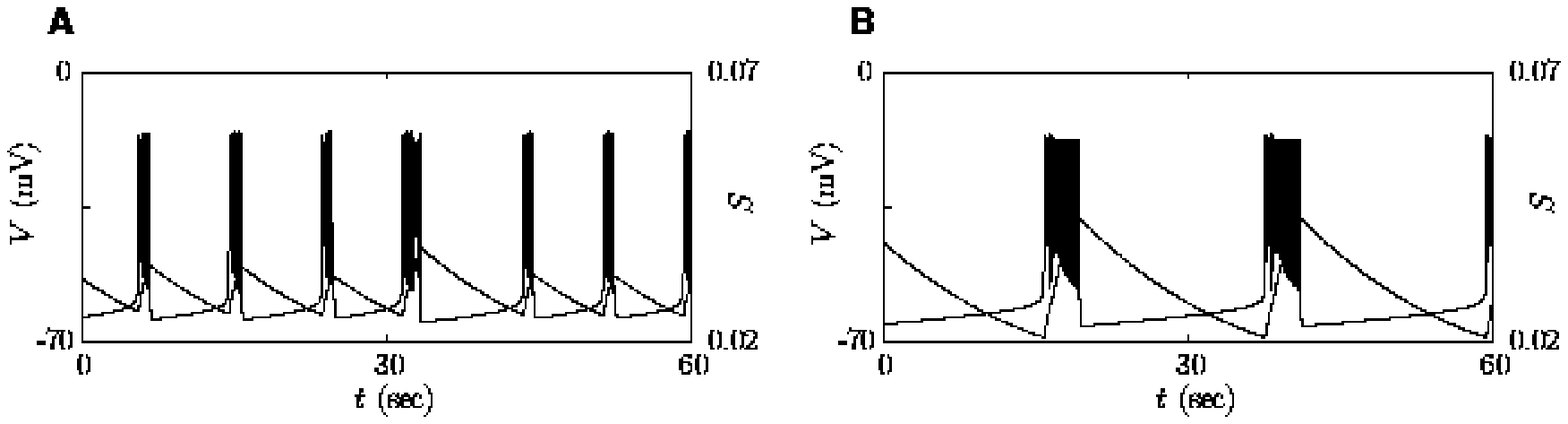}
      \caption{}
      \label{Fig:hetero}
   \end{center}
\end{figure}

\subsubsection*{Figure~\ref{Fig:hetero}.}
Bursting action potential induced by cell coupling,
with the cell-to-cell heterogeneity due to variations of the membrane capacitance $C_M$
and of the ATP-blockable K$^+$ channel conductance $g_{K(ATP)}$:
(\textit{A}) 20\,\% variation of $C_M$ (5.0\,pF, 6.3\,pF);
(\textit{B}) 10\,\% variation of $g_{K(ATP)}$ (1000\,pS, 1100\,pS).
Other parameter values have been taken from Table~I.

\clearpage

\clearpage

\clearpage

\clearpage

\clearpage

\clearpage

\clearpage

\clearpage

\clearpage

\clearpage

\end{document}